\documentclass[a4paper]{article}

\usepackage{INTERSPEECH2022}
\usepackage{hyperref}
\usepackage{float}
\usepackage{booktabs}
\usepackage[table]{xcolor}
\usepackage{graphicx}
\usepackage{array}
\usepackage{bbding}

\title{The Eloquence team submission for task 1 of MLC-SLM challenge}
\name{Lorenzo Concina$^1$, Jordi Luque $^2$, Alessio Brutti$^1$, 
Marco Matassoni$^1$, Yuchen Zhang$^3$ }

\address{
  $^1$Fondazione Bruno Kessler\\
  $^2$Telefónica Innovación Digital, Scientific Group\\
  $^3$University of Essex}
\email{\{lconcina, brutti, matasso\}@fbk.eu
jordi.luqueserrano@telefonica.com  yuchen.zhang@essex.ac.uk}

\begin{document}

\maketitle
\begin{abstract}

In this paper, we present our studies and experiments carried out for the task 1 of the Challenge and Workshop on Multilingual Conversational Speech Language Model (MLC-SLM), which focuses on advancing multilingual conversational speech recognition through the development of speech language models architectures. Given the increasing relevance of real-world conversational data for building robust Spoken Dialogue Systems, we explore three approaches to multilingual ASR. First,  we conduct an evaluation of the official baseline to better understand its strengths and limitations, by training two projectors (linear and qformer) with different foundation models. Second we leverage the SLAM-ASR framework to train a custom multilingual linear projector. Finally we investigate the role of contrastive learning and the extended conversational context in enhancing the robustness of recognition.
  
\end{abstract}

\section{Introduction}

Large Language Models have revolutionized natural language processing by enabling powerful and flexible systems for a wide range of text-based tasks. Extending these capabilities to spoken language is a next step toward building a multimodal human-computer interaction system. This has led to the emergence of Speech Language Models (SLMs)~\cite{surveySLM}, which aim to unify speech and language understanding within a single model architecture by extending LLMs to the audio modality. This approach aims to solve several tasks that were previously text-based in the audio domain. Among these are Automatic Speech Recognition, Spoken Language Understanding, and Emotion Recognition.

Traditionally, spoken language understanding systems have relied on cascaded pipelines, where ASR is followed by text-based language modeling and, in some cases, text-to-speech (TTS). However, this approach often suffers different issues: error propagation, lack of contextual integration between components, and the loss of paralinguistic cues and contextual dependencies due to the fact that the LLM only processes the transcription of the speech.

In this work, we describe our participation in the MLC-SLM Challenge to explore the development of multilingual conversational SLMs. Our focus is on building models that can operate directly on speech data in multiple languages, leveraging LLM-based architectures and the challenge data to improve recognition quality in complex, real-world dialogue settings. Additionally, through our participation, we aim to help disseminate the goals and ongoing research within the EU-funded  project Eloquence\footnote{https://eloquenceai.eu/}, which focuses on advancing multimodal, multilingual, and multi-task speech technologies~\cite{Shi2024MLSUPERB2B,Chen2024Floras5A}.

\section{Description of the system}
\begin{figure}[htbp]
  \centering
  \includegraphics[width=0.85\linewidth]{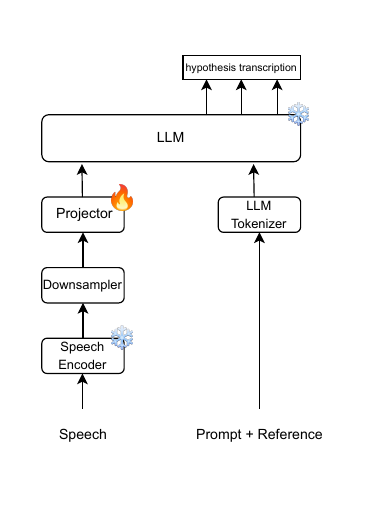}
  \caption{SLM pipeline}
  \label{fig:slm}
\end{figure}

In this study we faced the ASR task of the challenge with the Speech and Language Model (SLM)~\cite{SLM} approach, leveraging the power of an LLM to compute the transcription of the input audio, which is first passed to an audio encoder, then downsampled and finally projected by a linear projector to match the dimension of the input embeddings of the LLM. This approach is showed in figure ~\ref{fig:slm}. The LLM input is the concatenation of the processed input speech, the prompt (which undertake only the tokenization step) and, at training time, the reference transcription. 
In this system, the critical element is the projector, which must be trained while keeping the encoder and LLM frozen. Furthermore, it's possible to finetune the LLM together with the projector, or eventually use LoRA in case of an insufficient amount of training data. The projector and the LLM can be fine-tuned jointly, as demonstrated in our SLAM setup, or they can be fine-tuned in two separate stages, as implemented in the challenge pipeline. Our experiments indicate that the joint fine-tuning approach generally yields better performance.

In the challenge, we explored 3 different directions that are described in the following sections:
\begin{itemize}
    \item  we experimented with the challenge codebase provided as baseline\footnote{\href{https://github.com/mubingshen/MLC-SLM-Baseline}{MLC-SLM baseline}};
    \item we investigated the use of the SLAM-ASR\cite{ma2024embarrassingly} framework
    \item we explored the use of providing previous context as well as contrastive learning solutions. 
\end{itemize}

\subsection{Baseline Evaluation}

For comparison of the results, we experimented with the baseline code provided by the organizers. The MLC-SLM baseline is implemented using the Wenet~\cite{zhang22g_interspeech} Recognition Toolkit. The ASR model is composed of Whisper-large-v3 as speech encoder and Qwen2.5-7B or Llama3.1-8B as the backbone LLM. The training consists of two stages. In the first step a projector, composed of a convolutional layer with down-sampling factor 4 and a linear layer, is trained between the speech encoder and the LLM. In the following step, it simultaneously trains both the projector and LLM, the latter using Low-Rank Adaptation (LoRA), but starting from the projector trained in the first step. The table \ref{tab:example2} reports the result of the baseline submitted but using Whisper-large-v3 Turbo instead of the proposed baseline encoder. From the development experiments, the outcomes did not reveal a substantial difference when employed the two-stage training strategy. Consequently, we opted to submit additional trials wherein just the second training step was applied, that is, only training simultaneously the projector and the LLM.  Similar findings were observed in the context of switching the speech encoder, wherein using Whisper-large-v3 Turbo did not yield significant impact.

As reported in Table \ref{tab:example2}, we also compared several LLMs within the baseline Qwen2.5-7B: EuroLLM 1.7B and Salamandra with 2B and 7B parameters versions. For LoRA we follow the default settings provided by the baseline, with rank $r$ set to 16 and scaling factor $\alpha=8$. All speech encoders are feed with 128 log\_mel features and the training dataset is augmented using spectral masking augmentation and speed perturbation based on the SpecAugment~\cite{park19e_interspeech}, with default values as in the baseline provided recipe. The table also reports the results of a Querying Transformer (Q-Former) based projector. It is based on the Blip2QFormer implementation from the SLAM-ASR~\cite{ma2024embarrassingly} code. The Q-Former takes as input the entire output sequence of length 1280 from the Whisper encoder without down-sampling and applies two Q-Former hidden layers with a query length of 64.

\subsection{SLAM-ASR}
The SLAM-ASR~\cite{ma2024embarrassingly} framework is a multilingual Speech Language Model (SLM) that extends a pretrained Large Language Model (LLM) to process spoken input directly, following the architecture shown in figure ~\ref{fig:slm}. Although this framework presents yet unclear performances against issues such as being robust enough across different scenarios and speech
conditions, such as domain shifts and speech perturbations~\cite{SLAM-ASR-eval}, it is a powerful tool to experiment with different combinations of pre-trained foundational models and hyperparameters to achieve strong ASR capabilities. As the first line of experiments, we decided to leverage this tool to train a linear projector with different combinations of foundational models and hyperparameterers. Table ~\ref{tab:example} shows a list of the main experiments and the average WER computed in the challenge evaluation set. These experiments led to our best system, which is composed of Whisper Large V3 Turbo as audio encoder, a linear projector, and EuroLLM 1.7 ~\cite{EuroLLM} (frozen during training, but to which we apply LoRA). This experiment is the one appearing on the challenge public leaderboard, outperforming the baseline and placing thirteenth out of twenty-four submissions.
To further improve performances, we explored data augmentation on the train set, by applying several techniques, including Time Stretching, Pitch Shifting, Gaussian Noise and Clip Distortion. The reason is to increase the train data simulating a complex conversational scenario with low-quality audio. We first applied augmentation only to the languages that performed worse (Thai, Japanese and Portuguese) on our best systems, then to the full train set. Despite for some cases augmentation proved to be useful (for example, when using the instructed version of EuroLLM), we did not manage to improve the accuracy of the best performing model.

\subsubsection{Training Details}
All experiments presented in this section were trained using a consistent set of hyperparameters. The dataset was preprocessed with a mel spectrogram size of 128, ensuring uniform audio feature extraction across models. We used the Adam optimizer with a learning rate of 1e-4, training the projector for a total of 3 epochs.

To stabilize early training, we included a warmup phase of 1000 steps, followed by scheduled learning rate adjustments. During training, we used a batch size of 8 and a validation batch size of 2, balancing the GPU memory constraints with model performance evaluation.

As summarized in Table ~\ref{tab:example}, we tested several different LLM: EuroLLM 1.7B, EuroLLM 1.7 Instruct, EuroLLM 9B, Qwen 3 1.7B, Salamandra 2B, Salamandra 7B. EuroLLM 1.7B resulted to be the best performing choice both in terms of WER on the Eval set and resource usage during training and inference.

For models employing LoRA, the best performing one, which appears on the leaderbords, used the default parameters: rank $r$ set to 8 and scaling factor $\alpha=32$. We further explored different combinations of these hyperparameters using EuroLLM 1.7 Instruct. Table ~\ref{tab:Instruct+LoRA} summarizes the results on the Dev set (i.e. we did not submit those experiments). Note that using a larger rank is very beneficial; unfortunately we did not manage to test is with our best performing model (EuroLLM 1.7).

\begin{table*}[tb]
  \caption{Experimental results on the challenge evaluation set using SLAM-ASR}
  \label{tab:example}
  \centering
\begin{tabular}{l l l  l l l }
  \toprule
  \textbf{Audio Encoder} & 
  {\textbf{LLM}} & 
  {\textbf{LoRA}} & 
  
  \textbf{Projector} & 
   \textbf{Train data} &
  \textbf{Eval Set (WER)} \\
  \midrule
  \midrule
  Whisper l. v3 Trb &  EuroLLM 1.7B &  & Linear  & Full train set & 19.85\%  \\
    {\bf Whisper l. v3 Trb }&  {\bf EuroLLM 1.7B }& {\bf \checkmark }    &  {\bf Linear } &  {\bf Full train set} & {\bf 15.15\%}  \\
  Whisper l. v3 Trb &  EuroLLM 1.7B &\checkmark    & Linear  & Full train set + TH, JA, PT augment & 16.21\% \\
  Whisper l. v3 Trb &  EuroLLM 1.7B &\checkmark  & Linear    & Full train set + Full augment data  & 20.85\%  \\
  \hline
  Whisper l. v3 Trb &  EuroLLM 1.7B Instr. &\checkmark   & Linear& Full train set  & 19.11\% \\
  Whisper l. v3 Trb &  EuroLLM 1.7B Instr. &\checkmark   & Linear& Full train set + Full augment data & 16.86\%  \\
  \hline
  Whisper l. v3 Trb &  Qwen 3 1.7B &\checkmark    & Linear      & Full train set & 16.53\% \\
  \hline
  Whisper l. v3 Trb &  Salamandra 7B &\checkmark    & Linear      & Full train set & 18.64\%  \\
  Whisper l. v3 Trb &  Salamandra 2B &\checkmark   & Linear      & Full train set & 19.62\%  \\
  \hline
  Baseline - Whisper l. v3 &  Qwen2.5-7B &\checkmark    & Linear      & Full train set & 20.17\%  \\
\end{tabular}
\end{table*}

\begin{table*}[tb]
  \caption{Experimental results on the challenge evaluation and development sets using MLC-SLM baseline system based on Wenet. Results with $^{*}$ were not submitted and are given on the development set.}
  \label{tab:example2}
  \centering
\begin{tabular}{l l l  l l l }
  \toprule
  \textbf{Audio Encoder} & 
  {\textbf{LLM}} & 
  {\quad\textbf{Two stage train}} & 
  
  \textbf{Projector} & 
   \textbf{Epochs} &
  \textbf{Eval Set} \\
  & &\textbf{Projector + LoRA} & & &  \textbf{(WER)} \\
  \midrule
  
  \hline
  Whisper l. v3 Trb &  Qwen2.5-7B &\quad\quad\quad \checkmark + \checkmark   & Conv + Linear      & 6 + 5 & 22.47\%  \\
  Whisper l. v3 &  EuroLLM 1.7B Instr. & \quad\quad\quad -- + \checkmark    & Conv + Linear      & -- + 6 & 20.49\%  \\
  Whisper l. v3 &  Salamandra 2B Instr. & \quad\quad\quad -- + \checkmark    & Conv + Linear      & -- + 6 & 23.54\%  \\
  Whisper l. v3  &  Salamandra 7B Instr. &\quad\quad\quad -- + \checkmark    & Conv + Linear      & -- + 6 & 22.15\%  \\
  Whisper l. v3  &  Salamandra 7B Instr. &\quad\quad\quad-- + \checkmark    & Conv + Linear      & -- + 9 & $^{*}$21.13\%  \\
  Whisper l. v3 &  Salamandra 7B Instr. & \quad\quad\quad-- + \checkmark    & Q-Former + Linear  & -- + 6 & $^{*}$22.52\% \\
\bottomrule
\end{tabular}
\end{table*}

\begin{table}[h]
\centering
\caption{Experimental results on dev set with Whisper large v3 Turbo as encoder, EuroLLM 1.7B Instruct as LLM tuned with different LoRA parameters.}
\label{tab:Instruct+LoRA}
\begin{tabular}{ll}
\hline
\textbf{Experimental Setting} & \textbf{Dev Set (WER)} \\ \hline
LoRA ($\alpha = 8, r = 8$)  & 21.14\%      \\
LoRA ($\alpha = 8, r = 16$)      & 17.80\%           \\
LoRA ($\alpha = 16, r = 16$)  &  20.04\%          \\
LoRA ($\alpha = 32, r = 16$)  & 15.88\%          \\ \hline
\end{tabular}
\end{table}

\subsection{Contrastive and Context Learning}

We also extend our experiments by incorporating the last previous turn as context information for the ASR task. Prior studies show that contextual cues can assist in disambiguating speech content and improving transcription quality \cite{weng2020joint,masumura2021hierarchical,huang2024optimizing}. Building on this, we examine whether adding dialogue history and speaker tags can further enhance ASR results.

The procedure begins by reconstructing complete conversations from the training set using dialogue and speaker identifiers. This allows us to retrieve the most recent preceding utterance, hereafter referred to as the contextual turn, for each current speech segment. The contextual turn is then prepended to the input prompt, which is provided to the large language model (LLM) alongside the corresponding speech signal. The prompt is formatted as follows:

\begin{itemize}
    \item "[CONTEXT INFO]. Given the conversation history above between two speakers (O1 and O2), please transcribe the speech below. Speech: [AUDIO]. Transcription:"
\end{itemize}

To further leverage contextual signals, we incorporate a contrastive learning module that aligns the speech signal with its associated context. Specifically, each speech utterance and its context form a positive pair, while utterances paired with others' context serve as negative pairs. The objective of this contrastive learning module is to encourage the model to learn discriminative representations that bring positive pairs closer together in the embedding space, while pushing apart negative pairs. This approach is designed to enhance the model’s ability to jointly capture the semantic coherence between the speech content and its dialogue context. 
\subsubsection{Training Details}
For this experiment, we used EuroLLM 1.7B as the base large language model and train a linear projection head and a contrastive learning layer using the full training set. Evaluation is conducted on the development set. 
We used the Adam optimizer with a learning rate of 1e-4 and trained the model for a total of two epochs. Both the training and evaluation batch sizes were set to 16. For LoRA-based adaptation, we targeted the query and value projection matrices (q\_proj and v\_proj) within each self-attention layer. The LoRA configuration employed a rank of $r=8$ and a scaling factor of $\alpha=16$. All other model parameters were kept frozen during training.
The results, summarized in Table~\ref{tab:ctx}, demonstrate the effectiveness of incorporating both context information and contrastive learning in enhancing ASR performance. Interestingly, we observed that performance without LoRA was better than with LoRA. One possible explanation is that we only trained for two epochs, which may not have been sufficient for the model to converge when LoRA was activated.

\begin{table}[h]
\centering
\caption{Experimental results with context and contrastive learning}
\label{tab:ctx}
\resizebox{\linewidth}{!}{
\begin{tabular}{ll}
\hline
\textbf{Experimental Setting} & \textbf{Dev Set (WER)} \\ \hline
No context                                 & 19.85\%         \\
Context only (no contrastive learning)     & 18.92\%           \\
Context with contrastive learning (no LoRA) & 17.18\%          \\
Context with contrastive learning (with LoRA) & 18.62\%          \\ \hline
\end{tabular}
}
\end{table}

\section{Conclusions}
In this work, we explore three approaches to improve multilingual conversational ASR within the framework of the MLC-SLM Challenge. We first evaluated the official baseline by experimenting with different projectors and combinations of foundation models. We then leveraged the SLAM-ASR framework to systematically assess the impact of foundational model selection, training data augmentation, and model configurations on performance, identifying the most effective setup based on Whisper Large V3 Turbo and EuroLLM 1.7 with LoRA fine-tuning. Our best performing system ranked thirteenth on the public leaderboard, demonstrating the potential of lightweight, frozen LLM-based architectures combined with efficient projector training. Lastly, we introduced context-aware ASR modeling by incorporating preceding dialogue turns and applying contrastive learning, with the aim of improving transcription quality through semantic alignment and contextual grounding. These findings underscore the importance of both architectural choices and conversational context in advancing robust, multilingual SLMs. Future work will focus on deeper integration of dialogue structure, exploring cross-lingual knowledge transfer in low-resource settings and extend this architecture to a multi-task scenario.

\section{Acknowledgements}
This work has received funding from the European Union’s Horizon Europe research and innovation programme under the project ELOQUENCE (Grant Agreement No. 101135916).

\bibliographystyle{IEEEtran}

\bibliography{mybib}

@article{ma2024embarrassingly,
  title={An Embarrassingly Simple Approach for {LLM} with Strong {ASR} Capacity},
  author={Ma, Ziyang and Yang, Guanrou and Yang, Yifan and Gao, Zhifu and Wang, Jiaming and Du, Zhihao and Yu, Fan and Chen, Qian and Zheng, Siqi and Zhang, Shiliang and others},
  journal={arXiv preprint arXiv:2402.08846},
  year={2024}
}

@inproceedings{park19e_interspeech,
  title     = {SpecAugment: A Simple Data Augmentation Method for Automatic Speech Recognition},
  author    = {Daniel S. Park and William Chan and Yu Zhang and Chung-Cheng Chiu and Barret Zoph and Ekin D. Cubuk and Quoc V. Le},
  year      = {2019},
  booktitle = {Interspeech 2019},
  pages     = {2613--2617},
  doi       = {10.21437/Interspeech.2019-2680},
  issn      = {2958-1796},
}

@inproceedings{zhang22g_interspeech,
  title     = {WeNet 2.0: More Productive End-to-End Speech Recognition Toolkit},
  author    = {Binbin Zhang and Di Wu and Zhendong Peng and Xingchen Song and Zhuoyuan Yao and Hang Lv and Lei Xie and Chao Yang and Fuping Pan and Jianwei Niu},
  year      = {2022},
  booktitle = {Interspeech 2022},
  pages     = {1661--1665},
  doi       = {10.21437/Interspeech.2022-483},
  issn      = {2958-1796},
}

@article{surveySLM,
  title={A Survey on Speech Large Language Models},
  author={Jing, Peng and Yucheng, Wang and Yangui, Fang and Yu, Xi and Xu, Li and Xizhuo, Zhang and Kai, Yu},
  journal={arXiv preprint arXiv:2410.18908},
  year={2024}
}

@article{SLM,
  title={{SLM}: bridge the thin gap between speech and text foundation models},
  author={Mingqiu, Wang and Wei, Han and Izhak, Shafran and Zelin, Wu and Chung-Cheng, Chiu and Yuan, Cao amd Yongqiang, Wang and Nanxin, Chen and Yu, Zhang and Hagen, Soltau and Paul, Rubenstein and Lukas, Zilka and Dian, Yu and Zhong, Meng and Golan, Pundak and Nikhil, Siddhartha and Johan, Schalkwyk and Yonghui, Wu},
  journal={arXiv preprint arXiv:2310.00230 },
  year={2023}
}

@article{SLAM-ASR-eval,
  title={Performance Evaluation of {SLAM-ASR}: The Good, the Bad,
the Ugly, and the Way Forward},
  author={Shashi, Kumar and Iuliia, Thorbecke and Sergio, Burdisso and Esaú, Villatoro-Tello and Manjunath K E and Kadri, Hacioğlu and Pradeep, Rangappa and Petr, Motlicek and Aravind, Ganapathiraju and Andreas, Stolcke},
  journal={arXiv preprint arXiv:2411.03866 },
  year={2025}
}

@article{EuroLLM,
  title={{EUROLLM}
MULTILINGUAL LANGUAGE MODELS FOR EUROPE},
  author={Pedro, Martins and Patrick, Fernandes and João, Alves and Nuno, M. Guerreiro and Ricardo, Rei and Duarte, M. Alves and José, Pombal and Amin, Farajian and Manuel, Faysse and Mateusz, Klimaszewski and Pierre, Colombo and Barry, Haddow and José, G. C. de Souza and Alexandra, Birch and André, F. T. Martins},
  journal={arXiv preprint arXiv:2409.16235},
  year={2024}
}

@inproceedings{weng2020joint,
  title={Joint contextual modeling for asr correction and language understanding},
  author={Weng, Yue and Miryala, Sai Sumanth and Khatri, Chandra and Wang, Runze and Zheng, Huaixiu and Molino, Piero and Namazifar, Mahdi and Papangelis, Alexandros and Williams, Hugh and Bell, Franziska and others},
  booktitle={ICASSP 2020-2020 IEEE International Conference on Acoustics, Speech and Signal Processing (ICASSP)},
  pages={6349--6353},
  year={2020},
  organization={IEEE}
}

@inproceedings{masumura2021hierarchical,
  title={Hierarchical transformer-based large-context end-to-end asr with large-context knowledge distillation},
  author={Masumura, Ryo and Makishima, Naoki and Ihori, Mana and Takashima, Akihiko and Tanaka, Tomohiro and Orihashi, Shota},
  booktitle={ICASSP 2021-2021 IEEE International Conference on Acoustics, Speech and Signal Processing (ICASSP)},
  pages={5879--5883},
  year={2021},
  organization={IEEE}
}

@article{huang2024optimizing,
  title={Optimizing Large-Scale Context Retrieval for End-to-End {ASR}},
  author={Huang, Zhiqi and Caseiro, Diamantino and Joshi, Kandarp and Li, Christopher and Rondon, Pat and Wu, Zelin and Zadrazil, Petr and Zhou, Lillian},
  journal={Proc. Interspeech. ISCA},
  pages={4573--4577},
  year={2024}
}

@article{Shi2024MLSUPERB2B,
  title={ML-SUPERB 2.0: Benchmarking Multilingual Speech Models Across Modeling Constraints, Languages, and Datasets},
  author={Jiatong Shi and Shi Wang and William Chen and Martijn Bartelds and Bannihati Kumar Vanya and Jinchuan Tian and Xuankai Chang and Dan Jurafsky and Karen Livescu and Hung-yi Lee and Shinji Watanabe},
  journal={ArXiv},
  year={2024},
  volume={abs/2406.08641},
  }

@article{Chen2024Floras5A,
  title={Floras 50: A Massively Multilingual Multitask Benchmark for Long-Form Conversational Speech},
  author={William Chen and Brian Yan and Chih-Chen Chen and Shinji Watanabe},
  journal={2024 IEEE Spoken Language Technology Workshop (SLT)},
  year={2024},
  pages={891-898},
}


\end{document}